# A Review of Driver Drowsiness Detection Systems: Techniques, Advantages and Limitations

Ismail Nasri[1] , Mohammed Karrouchi[1] , Kamal Kassmi[1] , and Abdelhafid Messaoudi[2]

[1]Electrical Engineering and Maintenance laboratory, High School of Technology, Mohammed First University, Oujda, Morocco

[2]Energy, Embedded Systems and Information Processing laboratory, National School of Applied Sciences, Mohammed First University, Oujda, Morocco

*Abstract*—Driver Drowsiness is one of the most factors of road accidents, leading to severe injuries and deaths every year. Drowsiness means difficulty staying awake, which can lead to falling asleep. This paper introduces a literature review of driver drowsiness detection systems based on an analysis of physiological signals, facial features, and driving patterns. The paper also presents and details the recently proposed techniques for each class. We have also provided a comparative study of recently published works regarding the accuracy, reliability, hardware requirement, and intrusiveness. We have summarized and discussed the advantages and limitations of each class. As a result, each class of techniques has advantages and limitations. A hybrid system that combines two and more techniques will be efficient, robust, accurate, and used in real-time to take advantage of each technique.

## I. Introduction

According to the American National Highway Traffic Safety Administration (NHTSA) report in 2017, drowsy driving was responsible for 91,000 motor vehicle crashes, 795 deaths, and 4,111 fatalities in a motor vehicle crash between 2013 and 2017 [1]. The researchers suggest the prevalence of drowsy driving fatalities is more than 350% greater than reported [2].

Drowsiness refers to difficulty staying awake, even while performing activities. It is related to the biological wakefulness and sleeps linked to the circadian rhythm [3]. It is therefore not directly related to the accomplishment of a task. During a 24 hour cycle, the human body tends to sleep at certain times than at others. This is mainly the case during the night, between midnight and about 6 a.m., when the human body naturally turns towards sleep and vigilance decreases accordingly. There is no standard tool to measure the level of vigilance; the only solution is to observe the signs of hypervigilance emitted by the driver and analyze them. These signs can be divided into behavioral and physiological signs. Behavioral signs are manifested by abnormal behavior of the driver and represented by:

- A slowness of reaction,
- Inattention to the environment (road signs, obstacles, pedestrians, etc.),
- Coordination errors,
- An inability to maintain a fixed speed or trajectory.

The physiological signs appear as abnormal expressions, mainly on the driver's face and manifested by:

- Blinking,
- A stiff neck or back pain,
- Frequent yawning,
- Difficulty in keeping the eyes open and the head in a frontal position,
- Periods of micro-sleep.

When one of the signs appears, it is essential to take a break for at least 15 minutes before resuming driving. Unfortunately, drivers tend to overestimate their level of alertness and very often ignore these signs.

In 2014, the Belgian Institute for Road Safety (IBSR) conducted the first behavioral measurement of driver drowsiness in Belgium [4]. More than 2,500 respondents from a panel of 130,000 individuals participated in this study. At the beginning of the study, participants indicated whether they had driven a car in the last 24 hours. Immediately afterward, they were asked to recall one of the trips and answer questions about it as accurately as possible. Subjective sleepiness was measured using the Karolinska Sleepiness Scale (KSS) [5]. The study determined that 4.8% of trips involved a driver in KSS levels 6-9, for which drowsy driving is present (See Figure 1).

Most drowsy drivers report only low levels of drowsiness (Some signs of drowsiness = 3.3%). A driver makes only 0.5% of trips with "Sleepy, some effort to keep alert" and 0.1% with "Extremely Sleepy, fighting sleep". The results show that, in general, 4.8% of the trips made by drivers in Belgium are made by a driver showing signs of drowsiness.

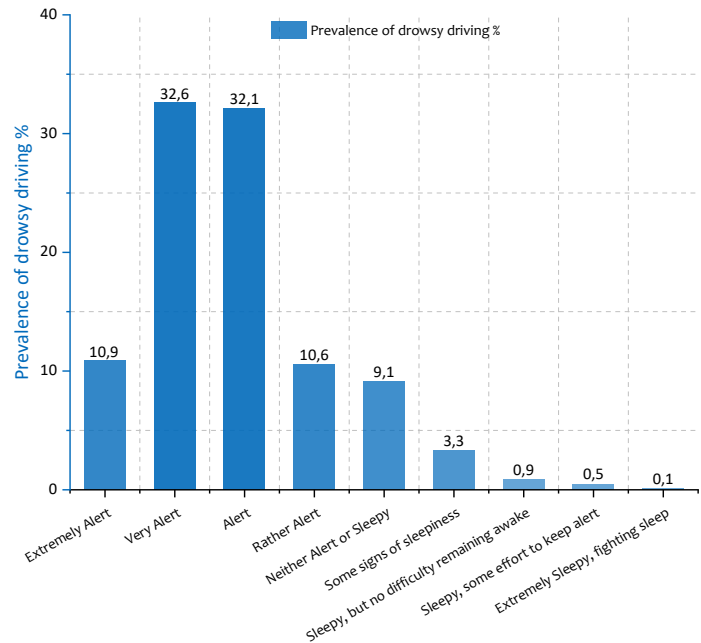

Fig. 1. Prevalence of driver drowsiness as measured with the Karolinska Sleepiness Scale (KSS)

Several techniques have been provided for detecting driver drowsiness. The researchers attempted to find the most reliable solution to detect driver drowsiness and design a system to prevent accidents caused by drowsy driving. In this paper, we have divided the techniques for drowsiness detection into three main classes according to the processed signal.

The main objective of this research paper is to introduce a detailed review of different driver drowsiness detection techniques, physiological signals analysis, facial features analysis, and driving patterns analysis to verify the effectiveness of each solution based on a set of performances such as accuracy, hardware requirement.

The remainder of this paper is organized as follows. Section 2 describes the reviewed methods used for driver drowsiness detection. The discussion and analysis are presented in Section 3. And the last section is dedicated to a conclusion along with future directions.

## 2. Methods

### 2.1. Physiological Signals Analysis

Physiological signal-based drowsiness detection systems use special sensors such as electroencephalogram (EEG) [6], electrocardiogram (ECG) [7], and electrooculogram (EoG) to measure the variation of signals such as brain activity or heart rate.

### 2.1.1. Electroencephalogram (EEG)

The electroencephalogram (EEG) measures the electrical activity of neurons through several electrodes. The frequency of brain activity in the EEG ranges from 1 Hz to 30 Hz and can be divided into four types according to frequency bands: delta δ (0-4 Hz), theta θ (4-8 Hz), alpha α (8-13 Hz) and beta β (13-20 Hz). The delta δ corresponds to the [0-4] Hz frequency band and is mainly present during a state of deep drowsiness; the theta θ corresponds to the [4-8] Hz frequency band and is associated with various psychological states involving a decrease in information processing. The alpha α corresponds to the [8-13] Hz frequency band, which is characteristic of relaxation. It is also very present when the eyes are closed. The Beta β corresponds to the frequency band [13-32] Hz. This activity is characteristic of wakefulness, active reflection and concentration. The Figure 2 below shows all components of the EEG signal, corresponding to the different frequency bands.

According to the physiological significance of the different EEG frequency bands, it is found that the EEG frequency bands that characterize decreased alertness and drowsiness are the θ and α band. The authors in [8], [9], [10], [11] also revealed that decreased alertness is characterized by frequency in the θ and α bands. The authors in [12] found that the vigilance decay is particularly apparent at α-band frequencies.

In most of the proposed systems for driver state inspection based on the study of alertness from EEG signals, the signals from the EEG sensor placed on the driver's head are recorded for several participants and labeled. Then, a supervised learning technique was used to conclude for each driver behavior the corresponding EEG signal. In [13], the authors note that it is difficult to obtain enough labeled EEG data to cover all states of alertness, and sometimes the labeled EEG data may not be reliable in practice. Therefore, they proposed a dynamic EEG-based clustering method for estimating alertness states. This method uses time series information to supervise the clustering of EEG data. Experimental results show that the method can correctly distinguish between the awake and drowsy states every 2 seconds using EEG, and can also distinguish two other intermediate states between awake and drowsy.

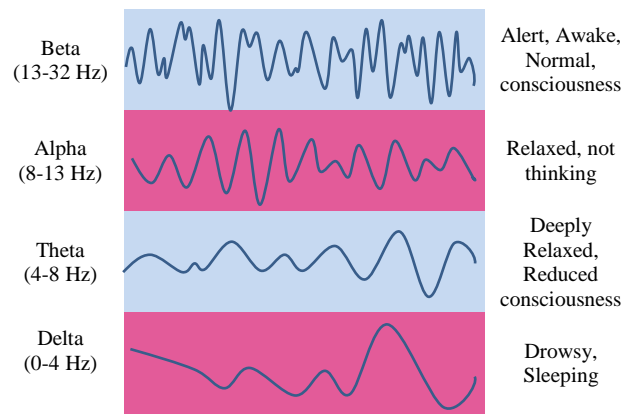

Fig. 2. EEG signals of different states.

### 2.1.2. Electrocardiogram (ECG)

The electrocardiogram (ECG) can be used to detect drowsiness by measuring the electrical activity of the heart. The degree of driver alertness can be determined by the variation in heart rate, a focused person will have a more regular heartbeat than a less focused individual. The ECG sensor (see Figure 4) records the electrical impulses that cause heart contractions, including LF (low frequency), VFH (very low frequency), HF (high frequency) signals and the LF/HF ratio. The wide variation in these signals reflects the state of the driver, such as alertness and drowsiness, at a distance from the heart, through the skin, using electrodes. The authors of [7] conclude that the ratio increases progressively during the transition from wakefulness to drowsiness, as shown in the Figure 3 below.

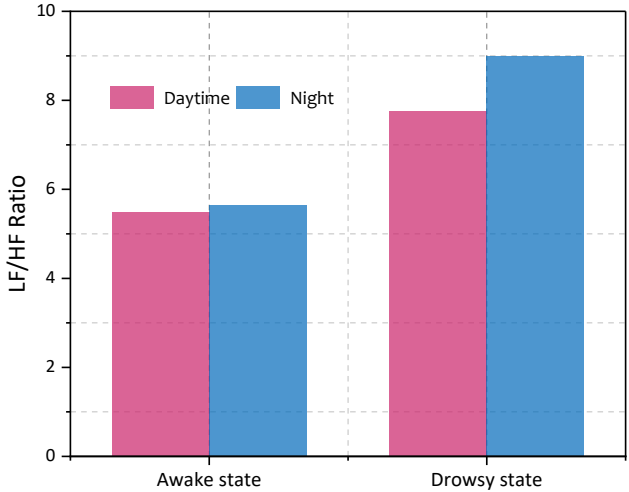

Fig. 3. LF/HF Ration in normal and drowsy state

The authors in [7], designed an ECG (electrocardiogram) sensor and a PPG (Photoplethysmogram) sensor to obtain physiological signals to monitor the driver's condition. The ECG and PPG signals are transmitted by sensors in the steering wheel to a base station connected to the server PC via a personal network for a practical test. An intelligent system then analyses and processes the HRV (Heart Rate Variability) signals, derived from the physiological signals, in the time and frequency domain and to assess the driver's drowsiness.

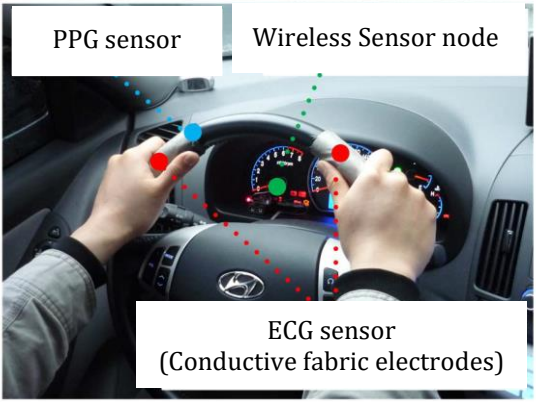

Fig. 4. ECG and PPG signal acquisition and transmission system on the steering wheel [7]

### 2.1.3. Electrooculogram (EoG)

Some researchers have used the electrooculogram EoG signal to identify driver drowsiness from eye movements [14], [15], [16]. The difference in electrical potential between the cornea and the retina generates an electric field that reflects the orientation of the eyes; this electric field is the measured EoG signal. The researchers studied horizontal eye movement by placing a disposable Ag-Cl electrode on the outside corner of each eye and a third electrode in the middle of the forehead for reference [15]. The electrodes were placed in such a way that the parameters Rapid Eye Movements (REM) and Slow Eye Movements

(SEM), which occur when a driver is awake and drowsy respectively, could be easily detected [17].

**2.1.4. Electromyogram (EMG)**

The electromyogram (EMG) is a recording of the electrical activity of active muscles. The EMG is a recording of the electrical activity of active muscles, which are low-intensity electrical currents. The use of this signal is based on the observation that during sleep, muscle activity decreases. For example, if the EMG amplitude decreases briefly and unexpectedly, the driver's state may correspond to a transition from wakefulness to drowsiness. If higher amplitude EMG discharges persist and are accompanied by increased alpha α activity (for >50% of the recording), the driver's state can be considered awake. The authors of [18] observed that the success rate of using a combination of EEG and EMG signals to detect drowsiness is higher than using either signal alone.

According to [6], [19], and [20], the physiological signal technique is the most accurate method for estimating the driver's state. However, their use remains limited due to the cost of the equipment and the fact that many sensors have to be attached to or in contact with the driver's body (see Figure 5), which is quite intrusive for the driver.

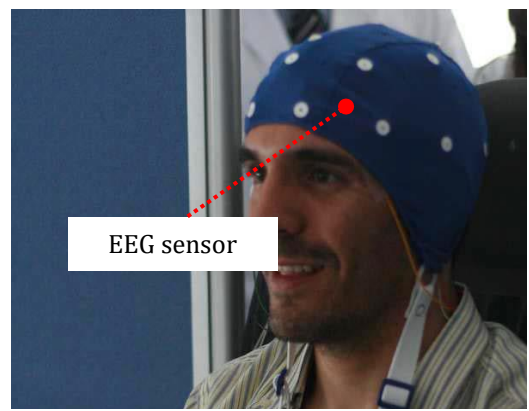

Fig. 5. ECG sensor placed on the driver's head [21]

## 2.2. Facial Features Analysis

The inspection of the driver's condition from physical signals relies mainly on the processing of the driver's video to measure the level of alertness reflected by his or her facial features. It has been observed that in case of fatigue, the driver exhibits certain visual behaviors that can be easily observed from changes in facial features such as yawning [22], eye closure duration [23], and head pose [24][25]. After locating the specific region of interest in the image, features such as PERCLOS, yawning frequency and head angle are extracted using feature extraction techniques. The percentage of eye closure as a function of time [53] and the frequency of eye closure are considered to be good indicators of drowsiness, as they allow the detection of micro-sleep periods [26] . Distracted driving has also been considered as one of the causes of road accidents; in this state, the driver may be concentrating on driving but is engaged in other activities while driving. The authors of [27], [28] and [29] developed a model for detecting driver distraction using a convolutional neural network (CNN) [30]. The proposed models were trained and tested using image datasets containing images of drivers in the most common activities, which lead to distraction while driving, including make-up, texting, calling, hands behind, etc. The main limitation of using an approach based on the analysis of driver facial features is the lighting. Normal cameras do not work well at night. In order to overcome this limitation, some researchers have used active illumination using an infrared light-emitting diode (LED) [23].

## 2.3. Driving Patterns Analysis

Monitoring vehicle behavior can indirectly reveal abnormal driver actions. Various parameters have been studied such as pedal force, gear change, Steering Wheel Movements (SWM), Standard Deviation of Lane Position (SDLP), etc. There are a limited number of vehicle brands that offer this kind of system as an option for some models. Volvo and Mercedes-Benz have introduced systems to determine the driver's status from the vehicle's behavior in the same period. In 2008, Volvo developed the first device in Europe, called Driver Alert Control (DAC) (DAC, 2008), to detect fatigue and alert the driver. Using sensors and a camera, the car alerts any driver who is visibly distracted or drowsy with an audible signal and a text message on the dashboard. A coffee cup icon is even displayed to prompt the driver to stop for a break. If the vehicle is drifting out of its lane without the driver having activated an indicator, the system will issue a warning signal.

### 2.3.1. Speed and Acceleration

In [31], the authors developed a solution to recognize five types of driving styles, such as normal, aggressive, drowsy, distracted and drunk driving, based on driving signals, including acceleration, speed, throttle position which controls the air intake of a spark ignition engine, gravity and revolutions per minute (RPM). Gravity and acceleration are obtained from a smartphone fixed in the vehicle. The other signals are collected from the vehicle using an on-board diagnostic connector (OBDII) [32] To take advantage of the deep convolutional neural network architecture on image processing, the collected signals are converted into images and passed through a trained CNN to classify an input image as one of five types of driving styles. Similarly, the authors of [33] sought to predict normal and aggressive driving style on the basis of several parameters, such as excessive speed, vigorous acceleration, number of lane changes, etc. The authors used two classifiers based on the results of their research. The authors used two classifiers based on neural networks: the Recurrent Neural Network (RNN) and the long short-term memory (LSTM) [34]. Both classification models were trained and evaluated using simulated data containing 17% aggressive driving and 83% normal driving. The simulated data was created and generated using the Luxembourg SUMO Traffic (LuST) scenario [35], which provides over 900 km of roads and 300,000 simulated vehicles. Some researchers have found that a state of hypovigilance can lead to high variability in driving speed [36].

### 2.3.2. Steering Wheel Movement (SWM)

The authors in [10], [37] used Steering Wheel Movement (SWM), measured with a steering angle sensor, to detect the driver's drowsiness level. An angle sensor mounted on the steering column measures the driver's steering behavior. In case of drowsiness, the number of micro-corrections on the steering wheel decreases compared to normal driving [38]. Similarly, the authors in [36] found that drivers with impaired alertness made fewer steering wheel reversals than normal drivers. To eliminate the effect of lane changes, the researchers considered only small steering wheel movements (between 0.5° and 5°), which are necessary to adjust the lateral position in the lane [10]. Thus, based on small SWMs, it is possible to determine the driver's drowsy state and give an alert if necessary. In a simulated environment, slight side winds pushing the car towards the right side of the road were added along a curved road to create variations in lateral position and force drivers to perform corrective SWMs [37]. Car manufacturers, such as Nissan and Renault, have adopted SWMs but they only work in very limited situations. This is because they can only work reliably in particular environments and are too dependent on the geometric characteristics of the road and, to a lesser extent, the kinetic characteristics of the vehicle.

### 2.3.3. Standard Deviation of Lateral Position (SDLP)

The measurement of the Standard Deviation of Lateral Position (SDLP) has been recognized as a reliable parameter for measuring driving impairment due to reduced alertness caused by drowsiness or sedative medication [39]. The SDLP represents the change in lateral position. It is an indicator of stability of path control. Its calculation requires the use of a front camera to measure the lateral position of the vehicle. The authors in [40] conducted an experiment to derive numerical statistics based on the SDLP and found that as the KSS [5] points increased, the SDLP (meters) also increased. For example, KSS scale points of 1, 5, 8 and 9 correspond to SDLP measurements of 0.19, 0.26, 0.36 and 0.47, respectively. The SDLP was calculated based on the average of 20 participants; however, with some drivers, the SDLP did not exceed 0.25 m even for KSS level 9. The authors note that driver fatigue would contribute to deterioration of lateral lane control (increased SLDP), while driver alertness would contribute to good lateral lane control performance (reduced SLDP). The SLDP-based technique is purely dependent on external factors such as road marking, weather and lighting conditions.

### 3. Discussion and Analysis

The physiological signals analysis technique is the most accurate method of all proposed techniques for driver drowsiness detection. However, their use remains limited because of the equipment's price, and that requires many sensors to be attached to the driver's body or put in contact with it, which is quite intrusive for the driver. On the contrary, methods based on driving patterns analysis are not intrusive. Indeed, some information directly related to driving activity is already available in the vehicle through the vehicle control network (Controller Area Network or CAN), such as accelerator and brake pedal pressure, steering wheel angle, etc., and do not require additional specific sensors. On the other hand, driving assistance systems, such as lane keeping assistance, already rely on information related to the vehicle's position in the lane. However, lane lines are not always visible. As a result, SDLP or the time to cross the line is not always available in real time. This is because SDLP requires robust detection of the trackside lines, which depending on the road conditions, is not always possible. It is also difficult to compare the numerical values and results presented by different studies as not all of them differentiate between bias and accuracy.

The main advantage of facial features analysis based-driver drowsiness these measurements, which justifies their important uses in the industrial applications of this research, is that they are easily accessible in a non-intrusive way as there is no need for sensors in contact with the participant. Therefore, they can be easily used in real life situations. Nevertheless, these measures rely on the fact that the acquisitions are based on image processing algorithms to detect the head, mouth, eyes and gaze. As a result, the quality of the estimation of the driver's state is highly dependent on the quality of the image processing. The latter is still difficult in the case of driving, especially for people wearing glasses, withstanding changes in light or in low light situations. In addition, there are a number of constraints related to the placement of the camera in the vehicle and the field of view covered by the lens. Table 1 provides the advantages and limitations of each class.

Table I. Advantages and limitations of each class

| Class | Hardware | Advantages | Limitations |
| --- | --- | --- | --- |
| Physiological Signals | Electroencephalogram (EEG) and Electrocardiogram (ECG) Sensors | Accurate, Reliable | Intrusive |
| Facial Features | External camera | Non-intrusive | Not possible in low brightness and all varying light conditions |
| Driving Patterns | Steering angle sensor for (SWM) External camera for (SDLP) OBD-2 for Acceleration, Speed, Gravity, and (RPM) | Non-intrusive | Unreliable, Less Accurate |

## 4. Conclusion

In this paper, we have reviewed the recent driver drowsiness detection techniques used in Advanced Driving Assistance System (ADAS) applications. The optimal use of these techniques in vehicles can prevent most accidents caused by drowsy driving every day. The presented and detailed techniques are divided into three classes according to the processed signals, including physiological signals, facial features, and driving patterns. The advantages and limitations for each class of techniques were also discussed in terms of accuracy, reliability, intrusiveness, and hardware requirement. As a result, each class of techniques has advantages and limitations. To take advantage of each technique, a hybrid system that combines two and more techniques will be efficient and able to be used in real-time.

As a result, the physiological signals analysis technique is the most accurate method for estimating driver drowsiness. However, their use remains limited because of the equipment's price and intrusiveness. We have also concluded that combining two drowsiness detection techniques can lead to significant performances and an efficient hybrid system used in real-time.

Further work will focus on designing and implementing an embedded system combining facial features and pattern analysis approaches to detect driver drowsiness, estimate the probability of accident risk and then alert the driver.